\documentclass{aa501}
\usepackage{graphicx}
\usepackage{natbib}
\begin{document}
   \title{The 3\,\mbox{$\mu$m} spectrum of R Doradus observed with the ISO-SWS\thanks{Based
on observations with ISO, an ESA project with instruments funded by ESA Member
States (especially the PI countries: France, Germany, the Netherlands and the
United Kingdom) and with the participation of ISAS and NASA. The SWS is a
joint project of SRON and MPE.}}


   \author{N. Ryde
          \inst{1}
      \and
          K. Eriksson
         \inst{2}
       }

   \offprints{N. Ryde}

   \institute{Department of Astronomy,
              University of Texas,
          Austin, TX 78712-1083,
          USA\\
              \email{ryde@astro.as.utexas.edu}
         \and
             Uppsala Astronomical Observatory,
         Box 515,
         SE-751 20 Uppsala,
         Sweden\\
             \email{kjell.eriksson@astro.uu.se}
         }

   \date{Received ; accepted }

   \abstract{We have modeled the $2.6$--$3.7\,\mbox{$\mu$m}$ spectrum
of the red semiregular variable R Doradus observed with the
Short-Wavelength Spectrometer on board the Infrared Space
Observatory. The wavelength resolution of the observations varies
between $\mathrm R\sim 2000$--$2500$. We have calculated a
synthetic spectrum using a hydrostatic model photosphere in
spherical geometry. The agreement between the synthetic spectrum
and the ISO observations is encouraging, especially in the
wavelength region of $2.8$--$3.7\,\mbox{$\mu$m}$, suggesting that
a hydrostatic model photosphere is adequate for the calculation of
synthetic spectra in the near infrared for this moderately varying
red giant star. However, an additional absorption component is
needed at $2.6$--$2.8\,\mbox{$\mu$m}$ and this discrepancy is
discussed.  The spectral signatures are dominated by water vapour
in the stellar photosphere, but several photospheric OH, CO, and
SiO features are also present. The effective temperature and
surface gravity derived for R Dor, based on the
$2.6$--$3.7\,\mbox{$\mu$m}$ ISO spectrum and the modeling of it
with a hydrostatic model photosphere, are $3000\pm 100\,\mbox{K}$
and $\log g = 0 \pm 1$ (cgs), respectively. The spectral region
observed is found to be temperature sensitive. The effective
temperature given here is slightly higher than those reported in
the literature. We also discuss possible reasons for this.
   }

   \authorrunning{Ryde \& Eriksson}
   \titlerunning{R Dor's $3\,\mbox{$\mu$m}$ SWS spectrum}
   \maketitle
%

\section{Introduction}


R Doradus is one of the brightest stars in the infrared. Its flux
distribution peaks in the 1\,\mbox{$\mu$m} range, the spectral
type being M8III\footnote{http://simbad.u-strasbg.fr/}. R Dor is
classified as a semi-regular variable [SRb; \citet{kholopov}], or
more precisely a red SRV [for a definition, see \citet{KH}], but
is closely related to the Mira variables in the sense that it
probably lies near the edge of a Mira instability strip
\citep{bedding}. R Dor has, however, a longer period than expected
for SRVs, but this fact could be due to irregularities in the
periods \citep{KH} or could be telling us that SRVs with long
periods represent a sub-group of Mira stars \citep{bedding:98}.

Semi-regular variables are most likely evolved stars on the
Asymptotic Giant Branch (AGB) in the HR diagram, see for example
\citet{habing}. AGB stars are low-initial-mass stars with
electron-degenerate, carbon-oxygen cores on their way from the
main sequence towards the white-dwarf/planetary-nebula phase.
Approximately 95\% of all stars will eventually pass this stage of
evolution \citep{habing}. In general, there seems to be an
evolutionary sequence on the AGB from the blue SRVs to the
long-period Mira stars, passing through the red SRV phase
\citep{KH,LH}. The blue SRVs are supposed to be on the early AGB,
the red SRVs are suggested to be experiencing the first thermal
pulses, and finally the Miras are associated with the thermal
pulses and are supposed to be pulsating in the fundamental mode,
showing large amplitudes in the visual brightness.

A key feature of AGB stars is their massive
but slow
stellar winds, creating an extensive
circumstellar envelope. AGB stars and their circumstellar
envelopes in general present spherical symmetry [see for example
\citet{ho:rev}].
In combination with the thermal pulses and the subsequent third
dredge-up, which drags up newly synthesised elements into the
atmosphere, the wind causes an enrichment of the interstellar
medium in elements synthesised in the AGB star, thus making these
stars important actors in the chemical evolution of galaxies [see
for example \citet{bg:96}]. R Dor has a mass-loss rate of
$7\times10^{-8}\,\mbox{M$_\odot$\,yr$^{-1}$}$, which is at the
lower end of what is typical for AGB stars, and a wind velocity of
$6\,\mbox{km\,s$^{-1}$}$ \citep{loup}. R Dor shows some
circumstellar dust emission [see discussions in \citet{bedding:98}
and \citet{KH}] and also molecular circumstellar emission from
regions close to the star [see for example \citet{ryde_CO2}].

Thanks to the Infrared Space Observatory (ISO), launched in 1995,
it has become possible to study the atmospheres of red giants in
detail at low and medium resolution, also at wavelengths obscured
by the Earth's atmosphere, for example the interesting region
between the K- and L-bands studied here.

Here we present the $2.60$--$3.66\,\mbox{$\mu$m}$
medium-resolution (R$\sim 2500$) spectrum of R Dor, observed by
the ISO-SWS, and model it by generating a synthetic spectrum from
the photosphere only. This results in a good agreement. The
photosphere is modeled
with the latest version of the {\sc marcs} model photosphere code,
generating hydrostatic photospheres. The study of infrared (IR)
spectra of red giants provides a test of the adequacy of
hydrostatic model photospheres, their input data and their
constraints, for the modeling of red giants.

In their unmasking of the warm molecular-envelopes surrounding AGB
stars \citet{tsuji_1997} model the $3 \,\mbox{$\mu$m}$ spectrum of
the M6 giant g Her and the M7 giant SW Vir observed by the
Short-Wavelength Spectrometer (SWS)
 on board ISO. \citet{markwick} model the
same wavelength region of ISO-SWS observations of the Mira star R
Cas (M7III). They suggest that the absorption spectrum originates
from the inner regions of the circumstellar envelope, and model it
with two circumstellar components, one warm and one cold.


\section{Observations}

   \begin{figure*}
   \centering
   \includegraphics{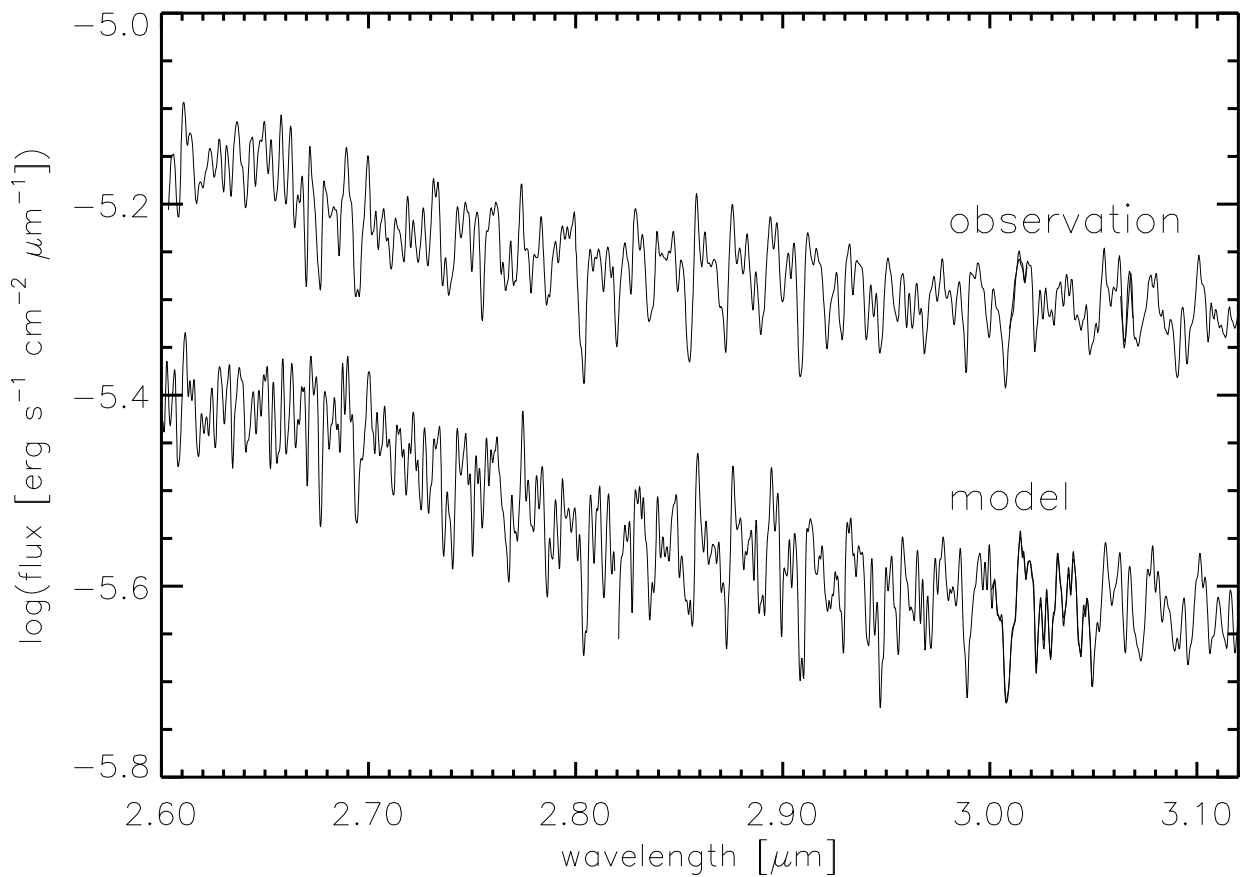}
   \caption{The first half of the observed spectrum of R Dor is shown as the upper spectrum. The
   modeled spectrum, which is shifted down for clarity, is shown below. The modeled spectrum is
   convolved with a Gaussian whose width matches the appropriate
   resolution of the bands observed.
   The abscissa
   scale shows the wavelengths in vacuum.}
              \label{fig1}%
    \end{figure*}
%

The spectrum of R Dor was observed with the SWS \citep{degraauw}
on board ISO \citep{kessler}. The spectrometer was used in the
grating scan mode (SWS06), which provides a resolution of ${\rm
R\sim 2000}$--$2500$, depending on the wavelength in the observed
region.

The 44 minutes of observations were performed on the 23$^{\mathrm
{rd}}$ June 1997 during ISO revolution number 585. The reductions
were made using the
pipeline, basic reduction package OLP (version 9.5) and the ISO
Spectral Analysis Package (ISAP version 2.0). The pipeline
processing of the data, such as the flux calibration, is described
in the ISO-SWS Handbook. The accuracy of the calibration of the
absolute flux is better than 7\% ($1\sigma$).

We also observed a low-resolution spectrum from
$2.4\,\mbox{$\mu$m}$ to $45\,\mbox{$\mu$m}$ of R Dor with the
ISO-SWS (in the SWS01 mode) on 27th June 1997 (orbit number 589).
Also here, the reductions were made using OLP version 9.5 and ISAP
version 2.0. Our SWS06 and SWS01 spectra of R Dor were thus
observed within 4 days of each other. The variation in the spectra
due to the periodic variations in the star can therefore safely be
ignored, R Dor having a period of 338 days \citep{kholopov}. The
SWS01 observation, in the wavelength region of our SWS06
observation, is composed of two sub-spectra; bands 1B and 1D.
These two sub-spectra could be merged into one spectrum directly.

Due to the way the spectrometer works in the SWS06 mode, our
region was observed in five spectral bands:
$2.60$--$3.02\,\mbox{$\mu$m}$, $3.02$--$3.08\,\mbox{$\mu$m}$,
$3.08$--$3.19\,\mbox{$\mu$m}$, $3.19$--$3.50\,\mbox{$\mu$m}$, and
$3.52$--$3.66\,\mbox{$\mu$m}$ [for details, see the ISO Handbook
\citep{handbook}]. There are mismatches in the fluxes between
these bands due to uncertainties in the detector gains
in every band. Furthermore, the overlap between the bands is
small. Therefore, in order to align the five bands, we used the
low-resolution SWS01 spectrum, convolved to
$\Delta\lambda=0.1\,\mbox{$\mu$m}$ or a resolution of R$\sim30$,
to outline the spectral shape of the region we
observed\label{backbone}.
In
this way we are able to scale the different bands of the SWS06
observation with correct factors in order to merge the bands into
one spectrum. The flux levels of the bands observed, lie within a
factor of 1.04 of each other. As a check, after having merged the
SWS06 sub-spectra, we also convolved the merged spectrum to
R$\sim30$, which resulted in the same shape as the SWS01
observation convolved to R$\sim30$. This indicates that we were
successful in merging the five sub-spectra into one spectrum over
the entire region and that we can rely on the overall shape of the
observed spectrum.

\begin{figure*}
   \centering
   \includegraphics{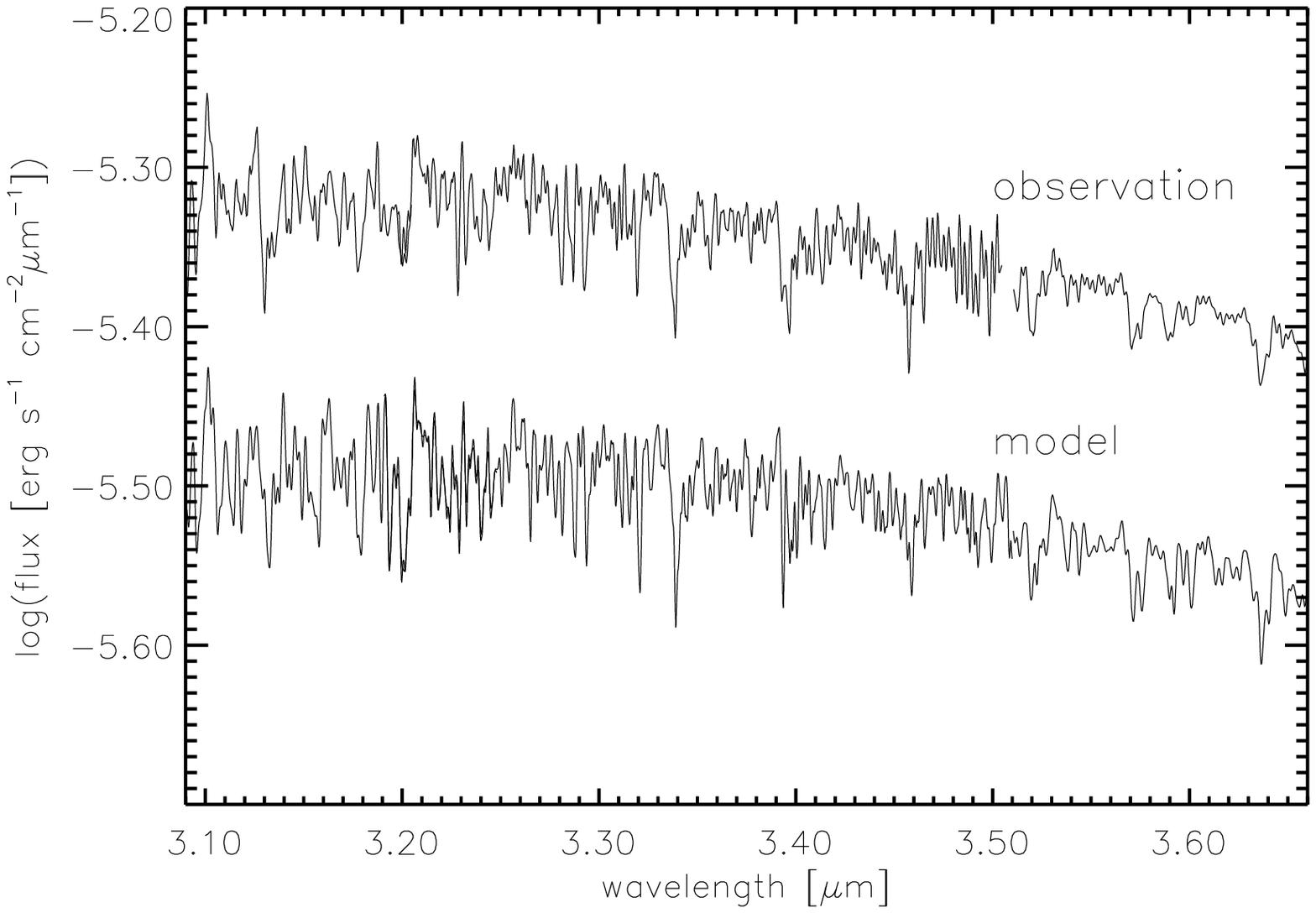}
   \caption{The second half of the observed wavelength range of R Dor. The observations are shown above and
the model spectrum is shifted down for clarity. The first half is shown in
   Fig. \ref{fig1}. The observed spectrum is a composite of several sub-spectra which
   sometimes overlap. However, at $3.51\,\mbox{$\mu$m}$ there is a gap in the
   observations. The various observed bands have different spectral resolutions.}
              \label{fig2}%
    \end{figure*}

In Figs. \ref{fig1} and \ref{fig2} we show the observed ISO
spectra.   The resolution of the five bands are $\mathrm R
=2\,300$, $2\,500$, $2\,000$, $2\,500$, and $2\,000$, respectively
\citep{handbook}.


\section{Model photospheres and the generation of synthetic spectra}

%

For the purpose of analysing our observations, synthetic spectra
have been generated \citep{IAU98:M} using model photospheres from
the new grid of spherically-symmetric model photospheres of M
giants, which is currently being calculated with the latest
version of the {\sc marcs} code. This version is the final major
update (of the code and its input data) in the suite of {\sc
marcs} model-photosphere programs first developed by
\citet{marcs:75} and further improved in several steps, e.g. by
\citet{plez:92}, \citet{jorg:92}, and \citet{BDP:93}.

These hydrostatic, spherical model photospheres are computed on
the assumptions of Local Thermodynamic Equilibrium (LTE),
homogeneity and the conservation of the total flux (radiative plus
convective; the convective flux being computed using the mixing
length formulation). The radiative field used in the model
generation, is calculated with absorption from atoms and molecules
by opacity sampling in approximately 84\,000
wavelength points over the entire, relevant wavelength range
considered for the star ($2300\,\mbox{\AA} $--$
20\,\mbox{$\mu$m}$).

Data on the absorption by atomic species are collected from the
VALD database \citep{VALD} and Kurucz (1995, private
communication). The opacity of CO, CN, CH, OH, NH, TiO, VO, ZrO,
H$_2$O, FeH, CaH, C$_2$, MgH, SiH, and SiO are included and
up-to-date dissociation energies and partition functions are used.
The continuous absorption as well as the new models will be fully
described in a series of forth-coming papers in A\&A (Gustafsson
et al., J\o rgensen et al., and Plez et al., all in preparation).
The new models were used by, among others, \citet{leen} in their
modeling of the K5III-giant $\alpha$ Tau.

Using the computed model photospheres we calculated synthetic
spectra by solving the radiative transfer at a high wavelength
resolution in a spherical geometry through the model photospheres.
Using extensive line lists (consisting of wavelengths, excitation
energies of the lower state of the transition, and line strengths
in the form of oscillator strengths), we produce synthetic spectra
of wavelength regions around the water vapour bands ($\nu_1$ and
$\nu_3$) at 2.60--3.66\,$\mu$m.
The line lists included in the generation of the synthetic spectra
are H$_2$O \citep{par} (including all lines stronger than given by
the condition $\mathrm {gf} \times 10^{-\chi \theta}/\max(\mathrm
{gf} \times 10^{-\chi \theta}) > 10^{-5}$, where
$\theta=5040/3050$ and $\chi$ is the excitation energy in eV,
leading to more than 2 million lines) , CO \citep{goor}, SiO
\citep{lang}, CH \citep{jorg}, CN (J\o rgensen \& Larsson, 1990,
\nocite{jorg_CN} and Plez 1998, private communications), OH
\citep{gold}, and C$_2$ (Querci et al. 1971 and J\o rgensen, 2001,
private communications\nocite{querci}). The accuracy and the
completeness of these line lists are discussed in
\citet{leen_phd}. We have also included CO$_2$ with data from the
Hitran database \citep{hitran,hitran2} and the Hitemp database
(Rothman et al. in preparation).

In the generation of synthetic spectra, we calculate the radiative
transfer for points in the spectrum separated by  $\Delta \lambda
\sim 1\,\mbox{km\,s$^{-1}$}$ (corresponding to a resolution of
$\lambda/\Delta \lambda \sim 330\,000$) even though the final
resolution is much less. With a microturbulence of $\xi\sim
2\,\mbox{km\,s$^{-1}$}$ in the model photosphere, this means that
we are sure of sampling all lines in our database in the
generation of the synthetic spectrum. This is an important point
since a statistical approach, by only taking fewer,
opacity-sampled points, will give an uncertainty in the
synthesised spectrum; There will be noticeable differences in the
synthesised spectra calculated with different sets of points,
unless one chooses the spacing between the points
smaller than the physical width
of the line broadening. This is especially important when dealing
with molecular bands, since the separation between lines differs
greatly with wavelength depending on whether the lines are close
to a band head or not. For example, by choosing random points in
the spectrum, at which the radiative transfer is calculated for
the synthetic spectrum, with a larger separation than we have
chosen, one will tend to overestimate the absorption at band heads
(or in regions with a high line density) and underestimate it far
from band heads (or in regions with a low line density).

The emergent model spectra are subsequently convolved with a
Gaussian in the same manner and with the same resolution as the
observed bands. This will allow a comparison with ISO
observations.

\section{Procedure and results}

In order to model the observations and to find a good fit, we have
calculated a grid of synthetic spectra, allowing the effective
temperature ($T_\mathrm{eff}$) and the surface gravity ($\log g$)
of the models they are based on, to vary by increments of 50 K and
0.5 dex, respectively.

First, the over-all spectral shape was matched. This was done by
convolving the observations and the models to a resolution of
$R\sim 60$. The wavelength coverage of $2.6$--$3.7\,\mbox{$\mu$m}$
is quite large and we find that especially the `low-frequency'
spectral shape, in particular the slope, at
$3.1$--$3.7\,\mbox{$\mu$m}$ is sensitive to the effective
temperature. This fact confines the possible range of effective
temperatures. Subsequently, the molecular features, both their
relative strengths and their amplitude are fitted visually. We
find that we are able to fit our observations well, both the
over-all shape and the molecular features, of the
$3\,\mbox{$\mu$m}$ region, for the phase at which the observations
were made, with a synthetic spectrum based on a model photosphere
of a temperature of $3000\,\mbox{K}$ and a surface gravity of
$\log g = 0 \mathrm{\,(cgs)}$, see Figs. \ref{fig1} and
\ref{fig2}. We point out that since the spectral region covered is
sufficiently large we can fit both the general shape, or slope, of
the spectrum as well as the molecular features themselves.
However, we are not able to fit the flux level at the beginning of
our observed region, at $2.6$--$3.8\,\mbox{$\mu$m}$. There is a
discrepancy here and this will be discussed in Sect. \ref{dis}. We
also performed a more formal $\chi^2$ fitting with the same
result. We determined the correct multiplicative factor to
multiply the modeled fluxes by (in order to account for the
distance to the star), to be able to compare them with the
observations, by minimising the $\chi^2$ for every model in our
grid. The parameters of the model giving the lowest overall
$\chi^2$ are also $3000\,\mbox{K}$ and $\log g=0$.



   \begin{figure*}
   \centering
      \includegraphics{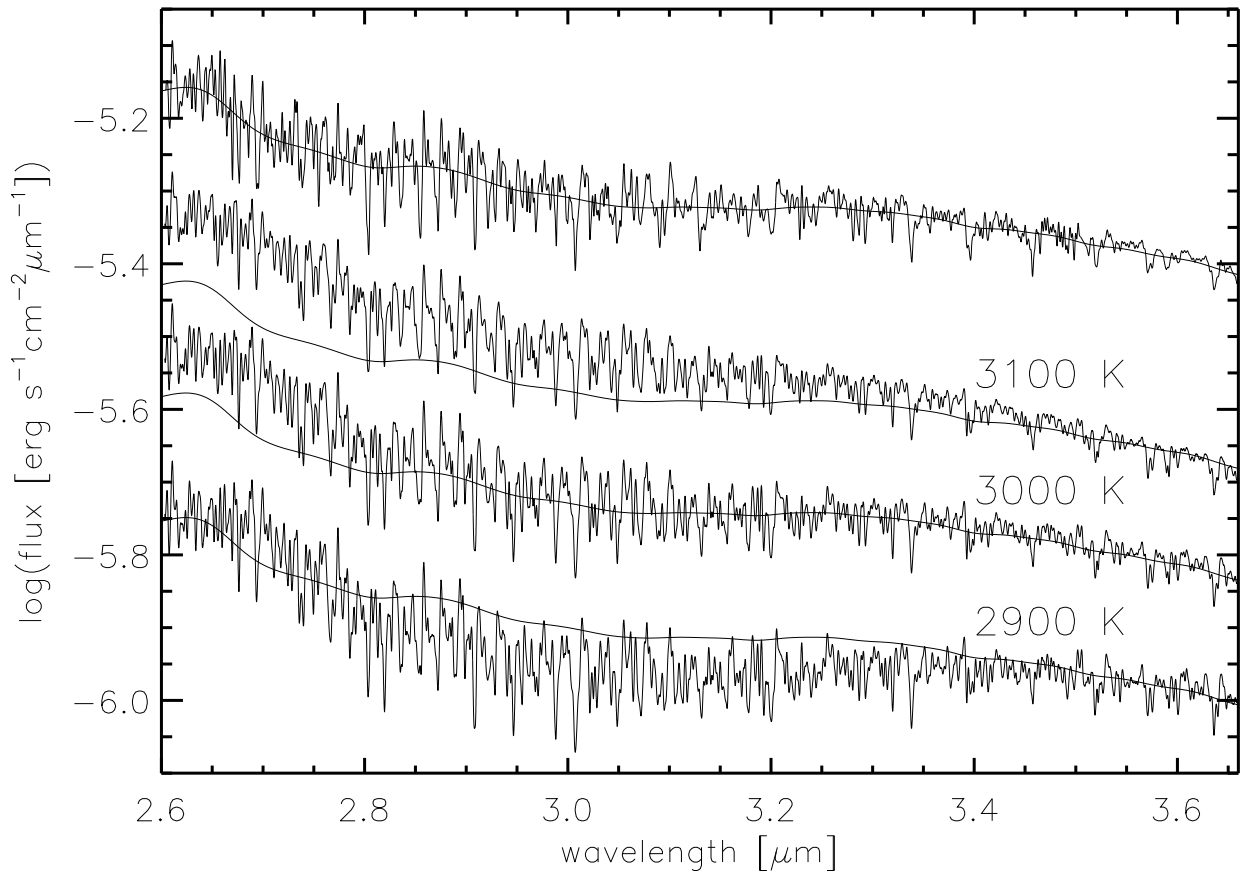}
      \caption{The observed spectrum is shown uppermost, and below
      are shown three
      synthetic spectra generated from model atmospheres of
      temperatures of (from top to bottom) $3100\,\mbox{K}$, $3000\,\mbox{K}$, and
      $2900\,\mbox{K}$ and of a surface gravity of $\log g = 0$.
      The SWS01 observation is shown to outline the observations
      and is repeated for the model spectra. These are arbitrarily
      normalised with the synthetic spectra at
      $3.5\,\mbox{$\mu$m}$.
              }
         \label{3100}
   \end{figure*}

Actually, a synthetic spectrum calculated on the basis of a model
photosphere with an effective temperature of $3100\,\mbox{K}$ fits
some of the molecular features
of the observed range better (though only marginally)  than does
the $3000\,\mbox{K}$ model, but this synthetic spectrum is then
too steep in the long-wavelength region and the overall strength
of the water bands is too weak, as is shown in Fig. \ref{3100}.
The uppermost spectrum in the figure shows the observations with
the SWS01 observation as a back bone to outline the spectrum. The
three spectra below are synthetic spectra based on model
atmospheres of temperatures of $3100\,\mbox{K}$, $3000\,\mbox{K}$,
and $2900\,\mbox{K}$.
The low-resolution SWS01 observation is repeated for the three
synthetic spectra.  The overall shape of the hottest model is too
steep, reflecting
weaker water bands. The reverse is true for the coldest model.
These facts may provide us with an estimate of the uncertainties
in the temperature. Thus, with a conservative estimate of the
uncertainties we arrive at a determination of the temperature of
$3000\pm 100\,\mbox{K}$ for R Doradus.

The surface gravity ($\log g$) is also a fundamental parameter,
and is varied together with the effective temperature in the
search for the best model, but the spectra are not as sensitive to
this parameter as to the effective temperature. For the grid of
synthetic spectra that we have generated, we chose an increment of
$\Delta \log g = 0.5 \mathrm{\,(cgs)}$. A change in surface
gravity changes the relative strengths and details of the
molecular features. We have chosen $\log g=0$ as our best model
but the uncertainty is quite large due to the low sensitivity of
this parameter. Based on the fits, we estimate the limits of the
surface gravity to be $\delta \log g = \pm 1$ (cgs). For a given
surface gravity the stellar mass does not greatly change the
spectral features or the spectral shape (typically by a few
percent for a change in mass by a factor of 3).

Also, the metallicity of the star has an influence on the
spectrum. This was also found by \citet{leen} in their study of
the red giant $\alpha$ Tau. We find that a decrease of the
metallicity by $0.15\,\mbox{dex}$ affects the absolute flux level
and the relative slope of the spectrum but only beyond
$3.3\,\mbox{$\mu$m}$. The molecular features hardly change. The
slope of the spectrum is changed in such away that for a given
effective temperature the slope decreases slightly for a lower
metallicity.

In addition to the observed spectrum, Figs. \ref{fig1} and
\ref{fig2} show our synthetic ones calculated with a hydrostatic
model photosphere of an effective temperature of 3\,000 K and a
surface gravity of $\log g = 0$. The stellar mass is assumed to be
1 solar mass and the chemical composition is assumed to be solar
\citep{feast}. For a metallicity of [Fe/H]$=-0.15$ the temperature
of the model that best fits the data is T$_\mathrm {
eff}=3050\,\mbox{K}$. In the figures of the spectra, the fluxes
are plotted on a logarithmic scale.
The uncertainties in the absolute ISO fluxes originate from
uncertainties in the gain factors of the different detectors (i.e.
observed bands) and not from the levels of dark current of the
bands. Allowing for these uncertainties and the distance to the
star, which has to be taken into account when comparing stellar
and model fluxes, the fluxes will differ only by a multiplicative
factor. A plot of the logarithm of the fluxes allows a direct
comparison with only an additive shift between them. The amplitude
of the spectral features can be compared directly.

   \begin{figure*}
   \centering
      \includegraphics{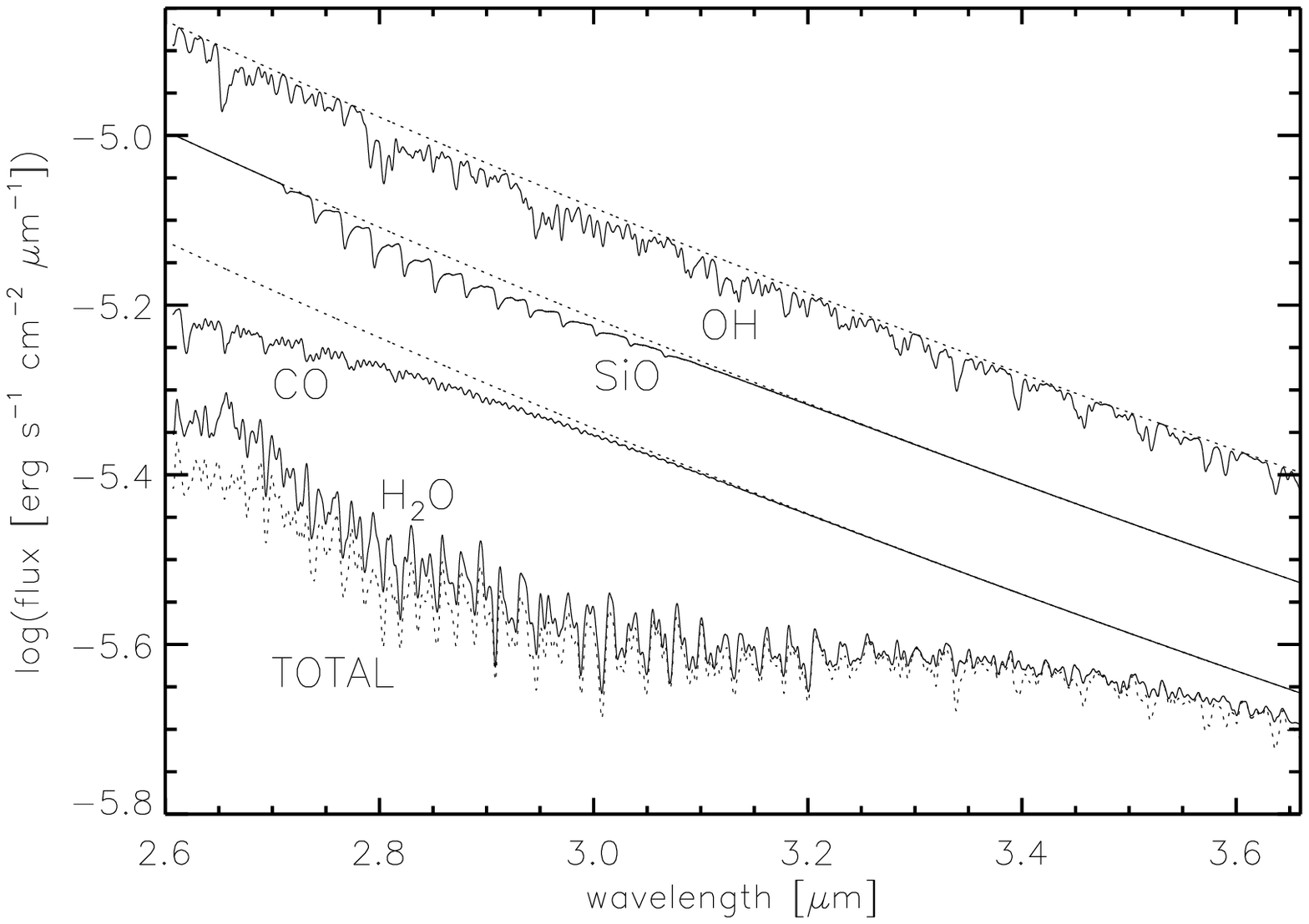}
      \caption{The total spectrum and the spectra of the
      contributing molecules
      are shown convolved with a Gaussian profile of a width
      corresponding to $300\,\mbox{km\,s$^{-1}$}$. This is done to give a better overview. The SiO and OH spectra
      are shifted up for clarity. The dashed, straight line represents the
      continuum. The continuum is also shown for the shifted SiO and OH
      spectra.
              }
         \label{bidrag}
   \end{figure*}

In Fig. \ref{bidrag}, spectra including the contributions of only
one molecule at a time, are shown.  The relative importance of the
H$_2$O, OH, CO, and SiO molecules contributing to the spectral
features in the observed region is displayed. The total synthetic
spectrum is also shown in the figure, and the SiO
and the OH spectra
are shifted for clarity. CO$_2$ does not contribute significantly;
the two vibration-rotational bands in this region are centered at
$2.69\,\mbox{$\mu$m}$ and $2.77\,\mbox{$\mu$m}$ \citep{herz},
but these are only 2\% deep as a maximum in our synthetic
spectrum. The other molecules included in the spectral synthesis,
i.e. CH, CN, and C$_2$, do not contribute to the spectrum at all.
All spectra are convolved with a Gaussian profile with a width of
$300\,\mbox{km\,s$^{-1}$}$, in order to provide a good overview of
the spectrum. As can be seen, the main features are due to water
vapour. It is the signatures of the $\Delta \nu_1=1$ and $\Delta
\nu_3=1$ modes of the rotational-vibrational bands of H$_2$O.
These modes are due to symmetric and antisymmetric stretching of
the water molecule. The first overtone vibration-rotational band
of the CO molecule starts contributing at $3.0\,\mbox{$\mu$m}$ and
increases toward shorter wavelengths. Several strong OH bands show
their signatures in the synthesised spectrum including all
molecules, especially at longer wavelengths
($3.3$--$3.7\,\mbox{$\mu$m}$). These OH signatures can also be
seen in the observed spectrum. The dashed line
shows the continuum of the spectrum.

The partial pressure of water vapour in the model photosphere
follows the gas pressure from the surface in to regions where the
temperature has reached approximately 4000 K, with values of
P(H$_2$O)/P$_\mathrm{g}$$\sim 3\times 10^{-3}$. Further in, the
partial pressure of water vapour decreases rapidly due to the
dissociation of the molecule.


\section{Discussion}
\label{dis}

The physical structure of AGB stars are certainly variable to
different extents. In order to model the atmospheres of these
variable stars in a proper physical manner one needs to take the
time dependence into account; this is increasingly important in
the infrared wavelength range where the light originates from the
outermost layers. The resulting dynamic effects could affect the
density and temperature structures of the outer photosphere and
therefore affect the line formation. For dynamic models of
variable stars see, among others, \citet{dorfi} and
\citet{bessel}. The partial pressures of molecules in the outer
photosphere may vary greatly over a pulsation period. However, the
fact that R Dor shows only a modest variability of $\delta\mathrm
V \sim 1.5\,\mbox{mag}$ (the AAVSO International Database; J. A.
Mattei 2001, private communication) suggests the feasibility of an
attempt to model the stellar photosphere of this red giant with a
hydrostatic model, as opposed to a dynamic model which takes into
account time variations of the photospheric structure.

\subsection{The hydrostatic modeling}

In the hydrostatic modeling of M giants, there are several
concerns to be addressed,
which could potentially  be important and explain discrepancies
between observations and their modeling.

For example, the main source of continuous opacity in red giants
in the IR spectral region is due to H$^-$ free-free processes, a
source of opacity that increases with wavelength. Therefore, IR
lines are formed far out in the photosphere, where the assumptions
made in the calculation of the model photosphere may be invalid.
The outer atmospheres of M giants are very tenuous and the
densities are low
in the line-forming regions. One should therefore also be aware of
the possibility of uncertainties due to NLTE effects affecting the
lines.

Furthermore, the input data regarding molecular lines (line
positions, excitation energies, and line strengths) and the
completeness of the data, i.e. whether all relevant molecular
bands are taken into account or not, are a concern. This is
especially important for water vapour since it dominates the
spectrum. Even bands far from the wavelength band considered, can
contribute significantly to the synthetic spectrum through a large
number of weak lines. Therefore, the inclusion of a relevant
number of lines, including weak ones, is important \citep[see
also][]{uffe_H2O}. \citet{jones} discuss the completeness of the
line list of water vapour by \citet{par}, which is used here. They
find a good match between their observations of M dwarfs and
synthetic spectra based on the Partridge \& Schwenke line list,
indicating its relevance for studies of M stars.

Finally, the relative extension ($d=\Delta r_\mathrm{atm}/R_*$) of
the photosphere, which is typically approximately $10\%$, means
that sphericity effects could be important both in the calculation
of the photospheric structure and in the radiative transfer, which
generates a synthetic spectrum. Indeed, we calculate our model
photospheres and synthetic spectra in spherical geometry as
opposed to plane-parallel geometry normally used for dwarf-star
models.

\begin{figure*}
   \centering
      \includegraphics{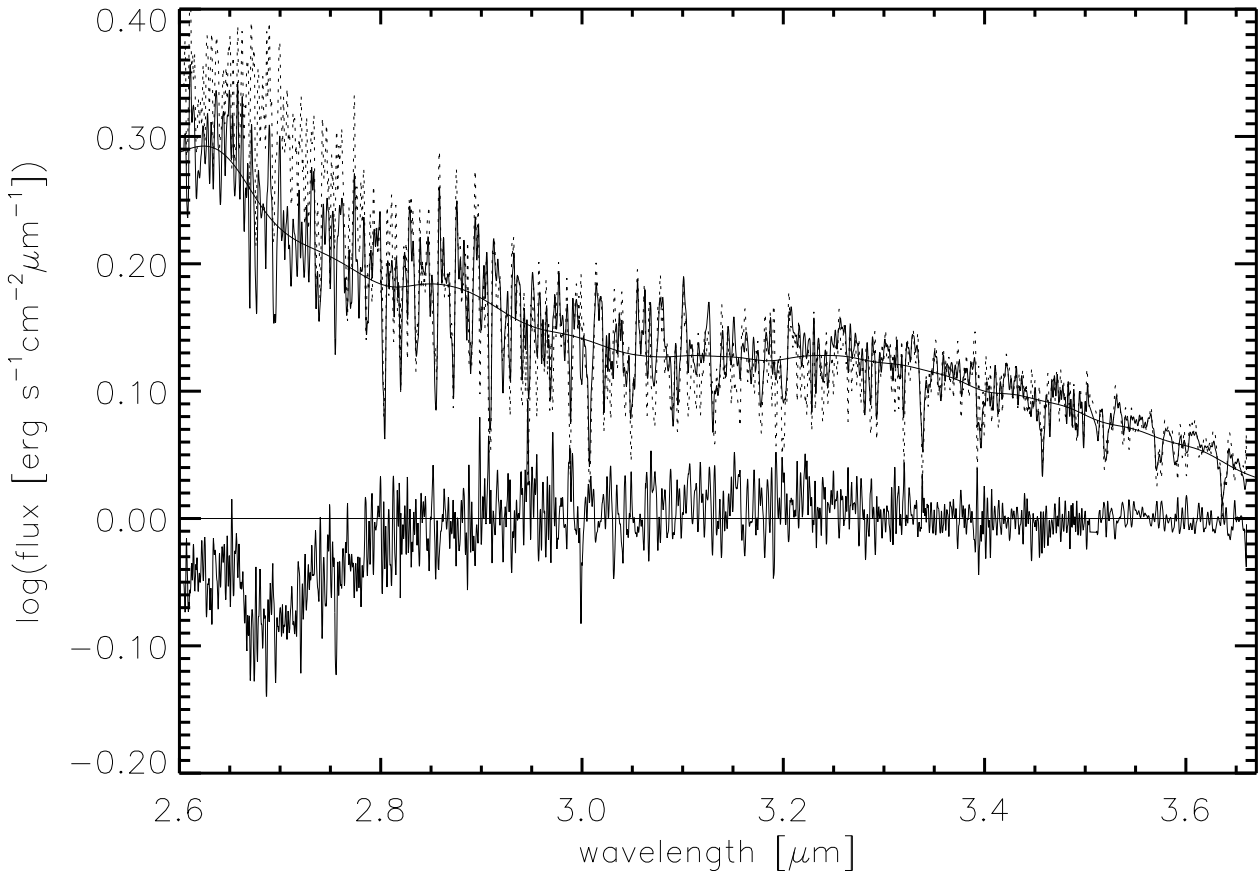}
      \caption{The observed (upper, full line) and synthetic spectra (upper, dotted
      line) are compared directly. In order to
      outline all the observed, medium-resolution spectra (SWS06) the
      low-resolution ISO observation (SWS01) is also shown. This observation is convolved to a resolution of
      $\Delta \lambda = 0.05\,\mbox{$\mu$m}$.
      The lower spectrum shows the ratio of the observed and the
      modeled spectra.
      The ordinata is given in logarithmic flux units arbitrarily scaled.
              }
         \label{tsuji}
   \end{figure*}

In spite of all the approximations made, the synthetic spectrum
presented here reproduces the ISO-SWS observations reasonably well
in the region of $2.8$--$3.66\,\mbox{$\mu$m}$, also keeping in
mind the uncertainties in the observational data and in the
reductions of the data. A direct comparison is presented in Fig.
\ref{tsuji}. The low-resolution SWS01 observation used to align
the different medium-resolution SWS06 observations (see page
\ref{backbone}) is also shown in the figure, as is the ratio of
the observed and the modeled spectra. The negative spike in the
ratio close to $3\,\mbox{$\mu$m}$ probably shows an inaccuracy in
the line list used. Thus, the use a hydrostatic model photosphere
for the synthesis of the photospheric spectrum of this spectral
region, seems to be quite adequate for the red, semi-regular
variable R Doradus. This is certainly not the case for the more
variable Mira stars [see for example \cite{aringer_leuven}].


\subsection{The discrepancy at $2.6$--$2.8\,\mbox{$\mu$m}$}

However, at the beginning of our interval, at
$2.6$--$2.8\,\mbox{$\mu$m}$, we find a discrepancy of the over-all
spectral shape, between our model and the ISO observations, see
Fig. \ref{tsuji}. We have not identified the cause of this, but we
will discuss a few points in connection with this discrepancy,
which reflect the approximations and shortcomings in the modeling.

First, the relevant absorption band of water vapour in the
wavelength region we are considering here has its maximum at
$2.65\,\mbox{$\mu$m}$ \citep{uffe_H2O,jones}. Therefore, one would
expect the lines at these wavelengths to be formed far out in the
photosphere and one might be concerned as to whether the extension
of the model photosphere is large enough to encompass this
line-forming region properly. In our models the water vapour lines
in the $2.6$--$2.8\,\mbox{$\mu$m}$ region are indeed formed
furthest out, with the strongest lines formed at an optical depth
of approximately $\log\tau_{\mathrm {Ross}}=-4.8$
($\log\tau_{\mathrm {Ross}}$ is the standard optical depth
calculated by using the Rosseland mean extinction coefficient).
However, our photospheric model is calculated out to
$\log\tau_{\mathrm {Ross}}=-5.6$ which means that the strongest of
the water vapour lines are formed within our photosphere and not
on the outer boundary of the model. As a check, we also calculated
a model extending to $\log\tau_{\mathrm {Ross}}=-6.2$, showing no
noticeable differences in the calculated spectra. Still, the
further out in the photosphere, the less likely are the
assumptions made in the calculations valid. Thus, the discrepancy
could have its origin in a too warm temperature structure in the
outermost layers. Furthermore, the model is calculated in LTE and
no line scattering is considered. Line scattering could yield a
larger absorption in the water lines. Neither should one rule out
dynamic effects here.

Second, some problems may exist concerning the input data;  the
line list calculations could be wrong just here, for instance due
to missing bands, but this is not likely [see the discussion in
\citet{jones}].


Third, there is certainly an extended atmosphere beyond the
photosphere of AGB stars, which could affect their spectra in the
infrared. The mass loss from the star creates a circumstellar
envelope which can extend far out. Also, it is seen in dynamic
model atmospheres of, for example, carbon-rich {\it Mira} stars,
that the photospheres could extend far out due to pulsations and
shocks \citep{dorfi}. (These basic features of dynamic model
photospheres are certainly valid also for oxygen-rich Mira stars.)
Densities at a few stellar radii from the star can be an order of
magnitude higher than the ones given by the equation of continuity
in the stellar wind.
Thus, the discrepancy could perhaps be explained by an additional
absorption from an extended atmosphere. A discrepancy similar to
the one we find between our modeled and observed spectra at
$2.6$--$2.8\,\mbox{$\mu$m}$, was also found by \citet{tsuji_1997,
tsuji_1998} in their analysis of the spectrum of the M7 giant SW
Vir. They attributed this to an excess absorption of warm
water-vapour of non-photospheric origin. In order to explain
several differences between models and infrared observations,
Tsuji and collaborators \citep[see for
example][]{tsuji_1997,tsuji_1998,tsuji_ny,tsuji_2000,yam_99} have
introduced the idea of a new, non-expanding, warm envelope
situated further out than the photosphere but distinct from the
cool, expanding circumstellar-envelope, at a distance of a few
stellar radii from the star. The origin of the envelope is neither
theoretically expected nor has it as yet received a theoretical
explanation, but seems to be a common feature of M supergiants and
M giants in general \citep{tsuji_1998,matsuura}. This new envelope
is shown to contain water vapour at temperatures of
$1000$--$2000\,\mbox{K}$ \citep{tsuji_1997}, resulting in
non-photospheric signatures in IR spectra of M giants. This view
was recently corroborated by ISO observations of the
$6.3\,\mbox{$\mu$m}$ bands of water vapour in early M giants,
stars not expected to show signatures of water vapour
\citep{tsuji_2001}. Thus, the discrepancy between our modeled and
observed spectra at $2.6$--$2.8\,\mbox{$\mu$m}$ could be a result
of an extra absorption component due to a non-expanding, warm
envelope. We are in the process of modeling such a region (Ryde et
al., in preparation). It is tempting to suggest in this scenario,
that a star with a sufficiently low mass-loss rate will not be
able to accelerate a wind from the stellar surface to velocities
above the velocity of escape, but instead accumulates material
"close" to the star where it could form a relatively dense, warm
layer including $\mathrm{H_2O}$.




\subsection{The derived effective temperature of R Dor}

Our derived effective temperature of $3000\pm 100\,\mbox{K}$,
which is based on the ISO spectrum of the
$2.60$--$3.66\,\mbox{$\mu$m}$ region, is somewhat higher than
values found in the literature.
For example, \citet{bedding} estimate an effective temperature of
$2740\pm190\,\mbox{K}$ from measurements of the angular diameter
and the apparent flux of R Doradus. As is discussed by these
authors, other indirect estimates of the effective temperature in
the literature yield even lower temperatures. We note that, given
the inferred radius of $R=370\pm50\,\mbox{$R_\odot$}$ of
\citet{bedding}, our temperature would yield a luminosity more
characteristic of a supergiant. However, we also note that
\cite{fluks}, who derive effective temperatures for all M-spectral
sub-types of the MK classification based on spectroscopic
observations of M giants in the solar neighbourhood, find an
$\mathrm T_{\mathrm{eff}}= 3126\,\mbox{K}$ for the M7 class. They
classify R Dor as an M7 giant and they derive an effective
temperature of $2890\,\mbox{K}$ for the M8 class.

Several possible reasons exist for the slightly higher temperature
deduced in this paper as compared to most of the literature
values. First, as was pointed out by \cite{bedding}, there may be
inadequacies in indirect methods of determining effective
temperatures from the colours of red stars. Second, \citet{jones},
in their fit of ISO spectra of M {\it dwarfs} with synthetic
spectra based on the line list of \citet{par}, also find effective
temperatures systematically higher than those found in the
literature, which are obviously based on several different
methods. They suggest that an explanation of this could be the
non-physical line splittings of the \cite{par} line list, which
lead to too strong water vapour transitions for a given
temperature. This would lead to a slightly higher effective
temperature for a synthetic fit of a stellar spectrum. Third, R
Dor is intrinsically variable, albeit with a small amplitude. As
has been discussed in, for example, \citet{aringer_PhD},
low-resolution ISO-SWS spectra of oxygen-rich, semi-regular
variables on the AGB can be modeled with a sequence of hydrostatic
models of different effective temperatures over the period of the
star. Thus, one would indeed expect slightly different effective
temperatures to be determined at different phases of the
pulsation.

\section{Conclusion}

We have attempted to synthesis an observed spectral region,
covering $1.06\,\mbox{$\mu$m}$ of the spectrum around
$3\,\mbox{$\mu$m}$ of the red giant R Doradus at a spectral
resolution of $\mathrm R \sim 2\,000$--$2\,500$. The synthesised
spectrum is based on a {\sc marcs} hydrostatic model-photosphere.
We are successful in reproducing  the
$2.80$--$3.66\,\mbox{$\mu$m}$ region, but encounter a discrepancy
in the beginning of the spectral region of our observations; $\sim
2.6$--$2.8\,\mbox{$\mu$m}$. We have discussed possible
explanations of this discrepancy, but have not identified the
cause.


The good agreement with the medium-resolution ISO-SWS06
observations, apart from the region with the strongest water
bands, suggests the adequacy of using a hydrostatic model
photosphere for this particular star, and the completeness and
correctness of our input data in the form of the molecular opacity
for the calculation of the spectrum. Given all the possible
failures of the model to correctly describe the physical picture
of the outer photospheres of red giants, the agreement is
promising.  It shows once again the accuracy and the strength of
the new {\sc marcs} code also in the infrared wavelength region.
Thus, the modeling of moderately varying red giants, such as
semi-regular variables, in the near-infrared region with
hydrostatic model photospheres may be a reasonable approximation.
Note, however, that for the very variable Mira stars in general,
hydrostatic model atmospheres certainly fail to reproduce observed
spectra [see, for example, \cite{aringer_leuven}]. Attention
should also always be given to the extended atmospheres
of these stars,
which could  affect the spectra and especially so in the infrared
wavelength region. This extended atmosphere is a possible reason
for the discrepancy in the $2.6-2.8\,\mbox{$\mu$m}$ region.


The spectral signatures in the spectrum presented here are mostly
due to photospheric water vapour but several photospheric CO and
OH bands are also identified. The spectral region is found to be
very temperature sensitive, leading us to an estimate of the
effective temperature of the star of $(3000\pm 100)\,\mbox{K}$.
The best-fit synthetic spectrum was based on a model photosphere
of a surface gravity of $\log g = 0 \pm 1$ (cgs). The temperature
found here is slightly higher than the ones reported in the
literature. A possible reason for this is a problem with the water
vapour line list of \citet{par}, as suggested by \citet{jones}.





\begin{acknowledgements}
We should like to thank Professors Bengt Gustafsson and David L.
Lambert for inspiration and enlightening discussions and the
referee for valuable comments and suggestions. This work was
supported by the P.E. Lindahl Foundation Fund of the Royal Swedish
Academy of Sciences
and
the Swedish Foundation for International Cooperation in Research
and Higher Education.
\end{acknowledgements}

\bibliographystyle{aabib99}

%


\end{document}